\documentclass[submission,copyright,creativecommons]{eptcs}
\providecommand{\event}{WORDS 2011} 
\usepackage{breakurl}             
\usepackage{pstricks}
\usepackage{pst-node}
\usepackage{amsmath}

\title{Word posets, with applications to Coxeter groups}
\author{Matthew J. Samuel
\institute{Department of Mathematics\\
Rutgers, the State University of New Jersey\\
Piscataway, New Jersey, USA}
\email{msamuel@math.rutgers.edu}
}
\def\titlerunning{Word posets}
\def\authorrunning{Matthew J. Samuel}
\begin{document}
\newtheorem{theorem}{Theorem}[section]
\newtheorem{conjecture}[theorem]{Conjecture}

\maketitle

\begin{abstract}
We discuss the theory of certain partially ordered sets that capture the structure of commutation classes of words in monoids. As a first application, it follows readily that counting words in commutation classes is \#P-complete. We then apply the partially ordered sets to Coxeter groups. Some results are a proof that enumerating the reduced words of elements of Coxeter groups is \#P-complete, a recursive formula for computing the number of commutation classes of reduced words, as well as stronger bounds on the maximum number of commutation classes than were previously known. This also allows us to improve the known bounds on the number of primitive sorting networks.
\end{abstract}

\section{Introduction}

If $S$ is a set of symbols, we recall that the free monoid $S^{\ast}$ generated by $S$ is the set of all words in $S$ of finite length. An arbitrary monoid generated by $S$ can be obtained from $S^{\ast}$ by declaring certain words to be equal, which determines an equivalence relation on $S^{\ast}$. For example, the relation $u=v$ indicates that whenever $u$ occurs in some word as a contiguous subsequence, the subsequence may be replaced by $v$ to obtain an equivalent word. If $a,b\in S$, a relation of the form $ab=ba$ is called a \emph{commutation relation}, and $a$ and $b$ are then said to \emph{commute}. Two words that are equivalent using only commutation relations are said to be in the same \emph{commutation class}. We will assume throughout that some set of monoid relations is fixed in advance. 

There is a method of representing commutation classes using partially ordered sets that captures their essential structure. We recall that a set $P$ together with a relation $\leq$ is a partially ordered set if $\leq$ is reflexive, transitive, and antisymmetric (meaning that if $x\leq y$ and $y\leq x$, then $x=y$); we will sometimes use $<$ to denote the corresponding antireflexive relation: $x<y$ if and only if $x\leq y$ and $x\neq y$. A finite partially ordered set $P$ together with a function $s:P\to S$ will be called a \emph{word poset} if the following conditions are satisfied for all $x,y\in P$:\\

\begin{itemize}
\item[(a) ] If $s(x)$ and $s(y)$ are equal or do not commute, then either $x\leq y$ or $y\leq x$, and\\

\item[(b) ] If $x<y$ and there is no $z\in P$ such that $x<z<y$, then we have either $s(x)=s(y)$, or otherwise $s(x)$ and $s(y)$ do not commute.\\
\end{itemize}

Two word posets $(P,s)$ and $(P',s')$ are said to be isomorphic, or in the same isomorphism class, if there is a bijective function $f:P\to P'$ such that $s'(f(x))=s(x)$ for all $x\in P$, and furthermore we require that $x\leq y$ if and only if $f(x)\leq f(y)$ for all $x,y\in P$. Then the first theorem on word posets is
\begin{theorem}
The isomorphism classes of word posets with $m$ elements are in bijective correspondence with the commutation classes of words of length $m$.
\end{theorem}

We illustrate with an example. Suppose we have $S=\{a,b,c,d\}$, which satisfy
$${ab}={ba}\mathrm{,}$$
$$cd=dc\mathrm{,}$$
and
$${ad}={da}\mathrm{.}$$
Then the word poset of $abcd$ is
$$\setlength{\arraycolsep}{1cm}
\psset{radius=1.5pt}
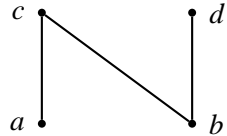
\begin{array}{cc}
\vspace{1cm}\Cnode*(0,0){C}&\Cnode*(0,0){D}\\
\Cnode*(0,0){A}&\Cnode*(0,0){B}
\end{array}
\ncline{A}{C}
\ncline{B}{C}
\ncline{B}{D}
\nput{180}{A}{{a}}
\nput{0}{B}{{b}}
\nput{180}{C}{{c}}
\nput{0}{D}{{d}}$$
Notice that if we read the labels in such a way that whenever two nodes are connected by a line, we read the bottom one first, we obtain a word in the same commutation class as $abcd$. This illustrates the following general result.

We recall that a linear extension of a partially ordered set $P$ with $m$ elements is a bijective function $e:P\to[m]=\{1,2,\ldots,m\}$ such that $e(x)\leq e(y)$ under the usual ordering on integers whenever $x\leq y$ in $P$. If $w$ is a word and $i$ is a positive integer, we will write $w_i$ to mean the $i$th symbol in $w$. Then
\begin{theorem}
If $(P,s)$ is a word poset, then the linear extensions of $P$ are in bijective correspondence with the words in the associated commutation class. If $e$ is a linear extension of $P$, then the corresponding word $w(e)$ is determined by $w(e)_i=s(e^{-1}(i))$.
\end{theorem}

The proof of this theorem is given as an exercise in Stanley \cite{stanley}.

Counting linear extensions of partially ordered sets is known to be a \#P-complete problem (see \cite{bw1991}). It follows from our work that the problem of counting words in commutation classes with arbitrary (finite) sets of monoid relations is polynomial time equivalent, and hence a first application is
\begin{theorem}
The problem of counting words in commutation classes of monoids is \#P-complete.
\end{theorem}
We have found further applications in the realm of Coxeter groups, which we now describe.

\section{Coxeter groups}

A Coxeter group generated by $S$ is a monoid determined by relations of the form $aa=1$ (where $1$ is the empty word) for all $a$ in $S$, as well as zero or more relations of the form $aba\cdots=bab\cdots$ for certain $a,b\in S$; one side of such a relation is an alternating word in $a,b$ of length $m\geq 3$ whose odd-index symbols are $a$ and even-index symbols are $b$, and the other side is the alternating word of the same length with odd-index symbols equal to $b$ and even-index symbols equal to $a$. Note that the commutation relation $ab=ba$ is in this form. The Coxeter groups generated by $S$ are in bijective correspondence with graphs on the vertex set $S$ whose edges are labeled either by $\infty$ or a positive integer greater than or equal to 3, known as \emph{Coxeter graphs} (see \cite{bb2005}). The lack of an edge between two vertices indicate that they commute, an integer label on an edge gives the length of the alternating word in a relation between the vertices, and a label of $\infty$ indicates that there is no relation.

Coxeter groups have applications in many areas of mathematics, but we will be interested mostly in the fact that they suggest interesting problems involving words. A word $w$ in a Coxeter group is said to be \emph{reduced} if there is no equivalent word of shorter length. Even if $S$ is infinite, an equivalence class of words in a Coxeter group will always contain only finitely many reduced words; in fact, it will contain at most $m!$, where $m$ is the length. Thus, a natural problem is to try to find formulas counting reduced words. Our work on word posets has resulted in a theorem that suggests that trying to find an efficient formula may not be a worthwhile endeavor:
\begin{theorem}
The problem of counting reduced words in Coxeter groups is \#P-complete.
\end{theorem}
There exist formulas counting reduced words in certain finite Coxeter groups, expressing the number of reduced words as a sum of certain easy-to-compute numbers (relating to Young tableaux, which are linear extensions of certain partially ordered sets) with coefficients. These formulas follow the work of Stanley in \cite{stanley1984}. However, computing the coefficients is itself a \#P-complete problem (see \cite{nh2006}), and the number of terms in the formula is superpolynomial, so the appeal of these formulas is not in their ease of use or efficiency. Rather, the formulas relate reduced words to Young tableaux, which are combinatorial objects that can be defined visually and are ubiquitous in other areas of algebraic combinatorics. Attempts to extend these formulas to completely general Coxeter groups have so far failed.

A formula for numbers of reduced words in general Coxeter groups in terms of linear extensions of partially ordered sets does arise naturally from our work. In the case of the finite Coxeter groups for which there are formulas in terms of Young tableaux, our formula is \emph{not}, in general, the same. If $P$ is a partially ordered set, denote by $E(P)$ the number of linear extensions of $P$. Also, we denote by $\mathrm{WP}(w)$ the set of reduced word posets (that is, word posets corresponding to commutation classes of reduced words) for $w$. Then our formula for the number of reduced words for $w$ is
$$\sum_{P\in \mathrm{WP}(w)}{E(P)}\mathrm{.}$$
The question then becomes: can we compute $\mathrm{WP}(w)$? 

In fact, the structure of Coxeter groups makes recursively constructing $\mathrm{WP}(w)$ relatively easy. The \emph{length} of $w$, denoted by $\ell(w)$, is the length of any reduced word for $w$. The set $D(w)$ of all elements $a$ of $S$ such that $\ell(aw)<\ell(w)$, called the \emph{left descent set} of $w$, can be determined in polynomial time. If we know $\mathrm{WP}(aw)$ for all $a\in D(w)$, then given $P\in \mathrm{WP}(aw)$ we can construct $P'\in \mathrm{WP}(w)$ by adjoining an element $x$ to $P$ such that $s(x)=a$ and $x<y$ for $y\in P$ if $s(y)$ does not commute with $a$ (then extending transitively). This leads to an inclusion-exclusion formula for the number of commutation classes of reduced words for $w$.

Denote by $C(w)=|\mathrm{WP}(w)|$ the number of reduced word posets for $w$. If $T\subset S$ is such that all elements of $T$ commute, then we will also denote by $T$ the product of all elements of $T$. Such a subset of $S$ will be called \emph{independent}, as it is an independent subset of the Coxeter graph. Then a recursive formula for $C(w)$ is given by
\begin{theorem}
We have
$$C(w)=\sum_{\substack{\emptyset\neq T\subset D(w)\\T\mbox{ independent}}}{(-1)^{|T|+1}C(Tw)}\mathrm{.}$$
\end{theorem}

For the purpose of enumerating words, we may assume that $|S|\leq\ell(w)$, because every reduced word for an element $w$ of a Coxeter group has the same set of distinct symbols as any other. If $|S|=n$, a trivial bound on $C(w)$ is 
$$C(w)\leq n^{\ell(w)}\mathrm{.}$$
This is trivial because $n^{\ell(w)}$ is in fact equal to the total number of words in $S^{\ast}$ of length $\ell(w)$. However, we have proved the following stronger bound.
\begin{theorem}
For any element $w$ in any Coxeter group with $\ell(w)>0$, we have
$$C(w)\leq \frac{2}{3}3^{\frac{1}{2}\ell(w)}\mathrm{.}$$
\end{theorem}
The best bound that was previously known was $3^{\ell(w)}$ for all $w$ in the symmetric group, or about $2.49^{\ell(w)}$ for $\ell(w)$ sufficiently large (see \cite{meng2010}). In any case, if it is always true that
$$C(w)\leq\alpha^{\ell(w)}$$
for some $\alpha$ and all $w$, then computations indicate that $\alpha>1.715$, and we conjecture that in fact we must have $\alpha\geq 3^{\frac{1}{2}}\approx 1.732$.

Let $M(k)$ denote the maximum number of commutation classes an element of length $k$ in any Coxeter group can have. We have that $M(0)=M(1)=M(2)=1$, $M(3)=M(4)=2$, $M(5)=3$, $M(6)=8$. We have shown that
$$\frac{1}{2.03669}\log{3}<\lim_{k\to\infty}{\frac{\log{M(k)}}{k}}\leq\frac{1}{2}\log{3}\mathrm{,}$$
and we conjecture that the upper bound is equal to the limit. We have also proved that
\begin{theorem}
An element of length $k$ with $M(k)$ commutation classes can always be found in some finite Coxeter group.
\end{theorem}
Thus, computing $M(k)$ can be done via a terminating algorithm.

\section{Sorting networks}

In the case that $w_0$ is the longest element of the finite Coxeter group of type $A_{n-1}$ for $n>0$ (i.e., the symmetric group $S_n$), $C(w_0)$ is equal to the number of primitive sorting networks on $n$ elements, as well as the number of rhombic tilings of a $2n$-gon (see \cite{armstrong2009}). Denote this number by $P(n)$. $P(n)$ for $n\leq 11$ had been computed and posted as sequence A006245 on the Online Encyclopedia of Integer Sequences before January $30^{\mathrm{th}}$, 2011. We were able to compute the $12^{\mathrm{th}}$ term using our formula, so that the terms posted as of the date of this writing are
$$1, 1, 2, 8, 62, 908, 24698, 1232944, 112018190, 18410581880, 5449192389984, 2894710651370536\mathrm{.}$$

Set $k_n=\frac{n(n-1)}{2}$. In \cite{wdmb2005}, it is stated that
$$0.23105\approx\frac{1}{3}\log{2}\leq\lim_{n\to\infty}{\frac{\log{P(n)}}{k_n}}\leq\log{2}\approx 0.69315\mathrm{.}$$
However, $P(n)\leq M(k_n)$ for all $n$, and we can prove that
$$\lim_{n\to\infty}{\frac{\log{P(n)}}{k_n}}\geq\frac{1}{k_m}\log{P(m)}$$
for any $m>1$. Thus,
$$0.53941\approx\frac{1}{66}\log{2894710651370536}\leq\lim_{n\to\infty}{\frac{\log{P(n)}}{k_n}}\leq\frac{1}{2}\log{3}\approx 0.54931\mathrm{,}$$
and we conjecture that the limit is equal to the upper bound.

\nocite{*}
\bibliographystyle{eptcs}
\bibliography{pragabstract}
\end{document}